\documentstyle[psfig]{mn}

\newif\ifAMStwofonts

\def\lapp{\ifmmode\stackrel{<}{_{\sim}}\else$\stackrel{<}{_{\sim}}$\fi}
\def\gapp{\ifmmode\stackrel{>}{_{\sim}}\else$\stackrel{>}{_{\sim}}$\fi}

\title[The magnetic field structure of SNR G328.4+0.2]
{The magnetic field structure of SNR G328.4+0.2\\ from polarimetric
observations at 19 GHz}
\author[Johnston et al.]
{Simon~Johnston$^1$, N.~M. McClure-Griffiths$^2$ \& B\"arbel Koribalski$^2$\\
$^1$School of Physics, University of Sydney, NSW 2006, Australia.\\
$^2$Australia Telescope National Facility, CSIRO, PO Box 76, Epping, NSW 1710, Australia.
}
\date{\today}
\pagerange{\pageref{firstpage}--\pageref{lastpage}}
\pubyear{2003}
\begin{document}
\maketitle
\label{firstpage}

\begin{abstract}
We report on the first polarimetric observations at 19~GHz made with the
upgraded Australia Telescope Compact Array. Observations were made
of the Galactic supernova remnant (SNR) G328.8+0.2. We find the
SNR has circular morphology with a strong central bar, similar to that
seen at lower frequencies. The SNR has high linear polarization throughout,
with fractional polarization in the bar up to 50 per cent.
The orientation of the magnetic field lines follow the filamentary structure
of the SNR. The magnetic field at the edge of the SNR is generally toroidal,
interspersed with radial fingers, likely caused by Rayleigh-Taylor
instabilities.
Although the SNR has been identified as Crab-like, we prefer an
interpretation in which the bar is a pulsar powered wind nebula with
the rest of the SNR consisting of the shell.
The proposed pulsar parameters make the SNR / pulsar system more like
SNR G11.2--0.3 than the Crab Nebula.
\end{abstract}


\section{Introduction}
Traditionally, supernova remnants (SNRs) have been classified into three
main classes, the most numerous of which are the shell SNRs. These are
characterised by synchrotron radio emission with a `shell' or
'barrel' like morphology. Typically, these shell SNRs have spectral indices
in the radio of $\alpha\sim$--0.7 (where $\alpha$ is defined in the sense
that $S_\nu \propto \nu^{\alpha}$).
However, the presence of an active, energetic
young pulsar near the centre of the SNR can give rise to composite SNRs.
In addition to the shell, these SNRs have a central component,
a synchrotron nebula, which is powered by a young and energetic pulsar. These
synchrotron nebulae generally have a rather flat spectral index
($\alpha\sim$0) and a high
degree of linear polarization. The third class of SNR, called Crab-like,
consist of pulsar powered nebulae without a detectable shell.
The Crab Nebula forms the archetype of this class, but
there are less than 10 Galactic SNRs with this classification.

SNR~G328.4+0.2 was first detected as a radio source in the early 1960s.
The most recent radio observations of the SNR were carried out by
Gaensler, Dickel \& Green (2000)\nocite{gdg00}. At 1.4 and 4.5~GHz,
the SNR has a diameter of $\sim$5 arcmin and a rather smooth appearance with
evidence for a central bar.
Caswell et al. (1975)\nocite{cmr+75}
derived a lower distance limit to the SNR of 17.5~kpc, based 
on H{\sc i} absorption measurements, a value confirmed by Gaensler et al. 
(2000).  The latter authors also showed that 
the SNR is highly polarized at 4.5~GHz and
that the rotation measure is --900 rad~m$^{-2}$. The spectral index
of the entire SNR is $\alpha\sim -0.12$.
Their conclusion was that G328.4+0.2 is an addition to the Crab-like
SNR class, i.e. it is a pulsar nebula with no visible shell.
By considering the energetics necessary to power the plerion,
Gaensler et al. (2000) showed that the
central pulsar must have a high spin-down energy, $\dot{E}\sim 10^{38}$
erg~s$^{-1}$, a short period and a relatively low magnetic field strength.
SNR~G328.4+0.2 was also observed in X-rays with the ASCA satellite by
Hughes, Slane and Plucinsky (2000)\nocite{hsp00}. The SNR was detected in the
hard X-ray band but the small angular diameter of
the SNR coupled with the large beam of ASCA meant that no detail was
observed. The emission appeared to be non-thermal and consistent with
other synchrotron emission nebulae.

Observations at high radio frequencies are not affected by
differential Faraday rotation and the small synthesized beam 
also minimises the effects of the tangling of the magnetic field along
the line of sight. This is particularly important in SNR G328.4+0.2
where the rotation measure is very high.
The Australia Telescope Compact Array (ATCA) has recently been upgraded
for operation at frequencies up to 100~GHz.
Each of the 22-m antennas has had surface upgrades and, during the past year,
all six antennas have been equipped with receivers operating over the
frequency range 16 to 24~GHz with receiver temperatures of $\sim$20--30 K.
The upgrade makes southern SNRs an immediate and compelling target for
polarization observations. In this paper we report on observations
of SNR G328.4+0.2 at 19~GHz. We derive the intrinsic magnetic field directions
and compare with magneto-hydro dynamic (MHD) simulations of SNR evolution.
We also speculate on the nature and classification of SNR G328.4+0.2.

\section{Observations}
Observations of SNR G328.4+0.2 were made with the ATCA, which
is an east-west synthesis telescope located
near Narrabri, NSW and consists of five 22-m antennas on a 3-km track with
a sixth antenna a further 3~km distant.
The ATCA can observe at two different frequencies simultaneously and is
capable of recording all four Stokes parameters at each frequency.

Our observations were obtained with two different array configurations,
750C and EW367, on 2003 June 17 and August 10. In these configurations, five
antennas were arranged to give maximum baseline lengths of 750\,m and 367\,m
respectively and a minimum baseline of 45~m in both cases.
Two different frequency bands were recorded, each with two
linear polarizations, centered at 18.752~GHz and 18.880~GHz. Each frequency
band was 128~MHz wide and contained 32 frequency channels each of width 4~MHz.
The central pointing position for SNR G328.4+0.2 was $\alpha,\delta$(J2000) =
$15^{\rm h}\,55^{\rm m}\,33^{\rm s}$, --53\degr\,17\arcmin\,00\arcsec.
The measured system temperatures
were $\sim$40--50~K for antenna elevations $\gapp$40\degr.
Because the primary beam at 18.8~GHz is only $\sim$1.8 arcmin and the SNR has
a diameter of $\sim$5 arcmin, we observed with seven separate pointings in a
hexagonal pattern separated by 1.1 arcmin. Each pointing was observed for
60~seconds, the pattern was observed twice before visiting the phase
calibrator PKS 1613--586 for 2--3 min, and this entire cycle was repeated for
12~hr in each array configuration. For bandpass calibration we used PKS 1921--293
(June) and PKS 0420--014 (August). The flux density of PKS 1921--293
varies substantially at high frequencies, and we therefore used PKS 1934--638
(flux density $\sim$1.15 Jy) as the primary flux calibrator.
Table~1 gives the calibrator fluxes we obtained with
respect to PKS 1934--638. We also observed the planets Mars and Uranus,
deriving scaling factors within $\sim$10\% of the adopted flux calibration.
Reference pointing was carried out every hour during the 
August observing run using the phase calibrator
PKS 1613--586 (flux density $\sim$3 Jy). The pointing correction was 
typically 15\arcsec\ to 20\arcsec\ with an rms of $\sim$8\arcsec.

The data were reduced in the {\sc miriad} software package using standard
procedures. After calibration, the {\em uv}-data (excluding the 6~km antenna)
were Fourier-transformed using multi-frequency synthesis of both bands and
`natural' weighting to produce Stokes $I$, $Q$, $U$ and $V$ 
mosaiced `dirty' images. We then performed maximum entropy deconvolution
of the polarization images using the task {\em pmosmem} \cite{sbd99}.
This task constrains $I$ to be positive and simultaneously
cleans the $Q$, $U$ and $V$ images.
The deconvolved images were restored with a synthesized beam of
$7\farcs7 \times 6\farcs1$. 
An image in linear polarization was computed by adding the $Q$
and $U$ images in quadrature.
Polarization vectors were computed when the linearly polarized flux
density exceeded 0.3 mJy\,beam$^{-1}$ ($\sim4\sigma$, where $\sigma =0.07$ 
mJy\,beam$^{-1}$ was obtained from the Stokes V image), the total flux 
density was in excess of 1.5 mJy\,beam$^{-1}$ and when the error 
in the vector was less than 10\degr.
\begin{table}
\begin{tabular}{lccc}
\multicolumn{1}{c}{Date} & \multicolumn{1}{c}{Source} &
\multicolumn{2}{c}{Flux Density (Jy)}\\
& & \multicolumn{1}{c}{18.752~GHz} & \multicolumn{1}{c}{18.880~GHz}\\
\hline
2003 June & 1934--638 & 1.154 & 1.146\\ 
     & 1921--293 & 13.46 & 13.49\\
     & 1613--586 & 3.18 & 3.17\\
     & 1253--055 & 16.16 & 16.19\\
     & 1334--127 & 3.96 & 3.97\\
2003 August & 1934--638 & 1.154 & 1.146\\
       & 1921--293 & 13.34 & 13.42\\
       & 1613--586 & 3.25 & 3.25 \\
       & 0420--014 & 14.23 & 14.27\\
\hline
\end{tabular}
\caption{Flux densities of the calibrators}
\end{table}

\section{Results and Discussion}
\subsection{Images of the SNR}
Figure~1 shows the total intensity image of G328.4+0.2. The expected flux
density of the SNR at 19~GHz, extrapolating from the Gaensler et al. (2000)
values at lower frequencies, is $\sim$10~Jy whereas we measure 
only 4--5 Jy. This is almost
certainly because the interferometer is filtering out the emission on scales
larger than resolved by our shortest baseline, i.e. 1.6 arcmin. 

The SNR is almost perfectly circular with a diameter of 5 arcmin. The extent
of the SNR is identical to that seen at lower radio frequencies.
The 19~GHz image has a more mottled appearance 
than at low frequencies but this is caused by the lack of short spacings
in the high frequency image. Spatial filtering of the low frequency image
presented in Gaensler et al. (2000) results in an image that is virtually
identical to the one shown here, including the filamentary structure.
As seen in Figure~1, the central bar is clearly delineated and 
has an extent of 7$\times$2~pc (assuming a distance of 17~kpc).
It runs east-west and is slightly offset
to the south from the geometric centre of the SNR. 
The circular morphology and the enhanced emission from the interior
is similar to that of G11.2--0.3 \cite{ggts88}.

The SNR is highly polarized throughout, with fractional polarization of up to
50 per cent in many regions. At this frequency, depolarization due to
either high rotation measures across the band or large variations in the
rotation measure from the back to the front of the SNR are unimportant and hence
the polarization that we observe is significantly higher 
than at lower frequencies. Furthermore, even the high rotation measure
of 1000~rad~m$^{-2}$ (Gaensler et al. 2000) imposes a turn of only
$\sim$10\degr\ on the position angle of the polarized radiation giving
us confidence about the derived magnetic field vectors. 
Figure~2 shows the total intensity overlaid with vectors
showing the direction of the magnetic field.
The magnetic field clearly follows the direction of the
filamentary structures in the SNR, with the twists and turns showing equally
well in both total intensity and magnetic field direction.
At the boundary between the SNR and the
ambient medium, the magnetic field is clearly toroidal (especially in the
north) with evidence of radial structure, e.g. in the south-east, where
the remnant departs from its circular appearance.
We note that we are not sensitive to the large scale magnetic field
structure (if it exists) due to the lack of short spacings in the image but
we can be confident in the magnetic field vectors tracing the small
scale structures.

\subsection{The bar in G328.4+0.2}
Fig.~3 shows an enlarged view of the central bar at a resolution of 2\arcsec.
The region with the
highest linearly polarized intensity is at the centre of the bar
and offset to the east from the peak in total intensity.
The magnetic field runs parallel to the major axis of the bar except
where it appears to be crossed by a foreground filament. The bar
shows signs of bifurcation, especially in the west.
The nature of the bar, in size, morphology and magnetic field orientation
is remarkably similar to the bar in the composite SNR G326.3--1.8 \cite{dms00}.
In that SNR, the bar is also elongated, shows signs of bifurcation at
the ends and has a magnetic field which runs parallel to the major axis.
If one assumes that a pulsar is located in the centre of the bar and
that it originated in the centre of the SNR at the time of the explosion
then in both SNR G328.4+0.2 and SNR G326.3--1.8 the orientation of the major
axis of the bar is perpendicular to the inferred velocity vector of the
pulsar. A second example of a plerion with a similar structure is
SNR~3C58. Although that SNR shows no shell, the plerion again has
a similar size, morphology \cite{bkw01,ra88} and polarization \cite{ww76}
to SNR~G328.4+0.2.
\begin{figure}
  \centerline{\psfig{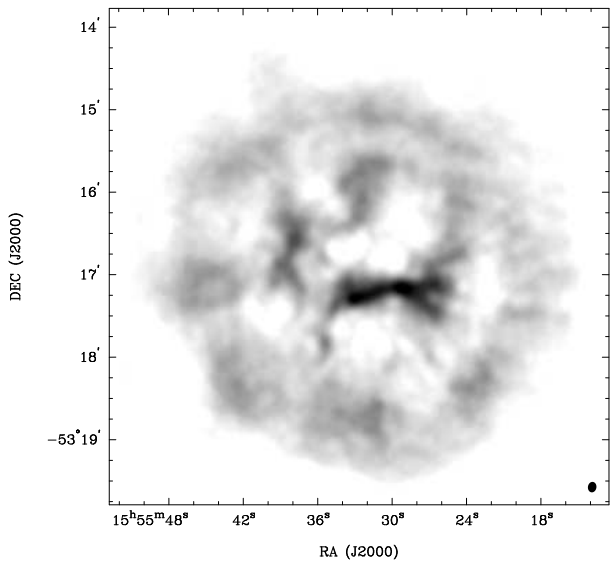}}
  \caption{Total intensity map of G328.4+0.2. The grey scale runs from
1.5 to 13 mJy\,beam$^{-1}$. The synthesized beam is shown in the bottom
left corner.}
\end{figure}
\begin{figure}
  \centerline{\psfig{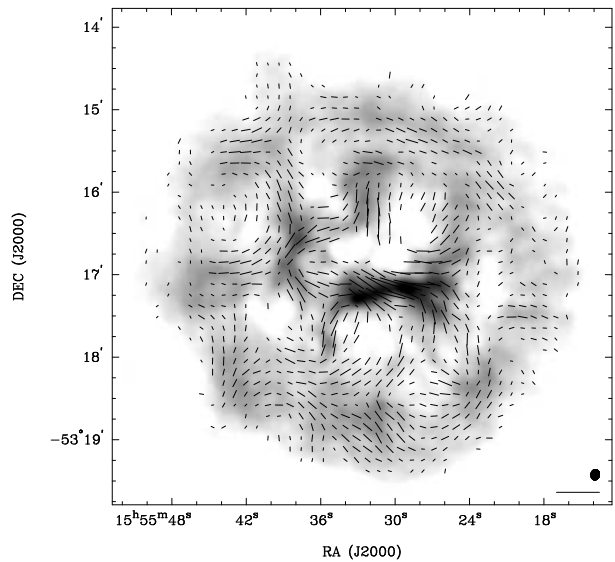}}
  \caption{Total intensity map of G328.4+0.2 overlaid with vectors showing
the orientation of the magnetic field (i.e. they are perpendicular to
the measured polarization angles).  The grey scale is as for Fig.~1.
The length of the vectors is proportional to the linearly polarized
intensity. The synthesized beam is shown in the bottom right corner along
with a vector of length 5.5 mJy\,beam$^{-1}$.}
\end{figure}

\subsection{Magnetic field structure}
\begin{figure}
  \centerline{\psfig{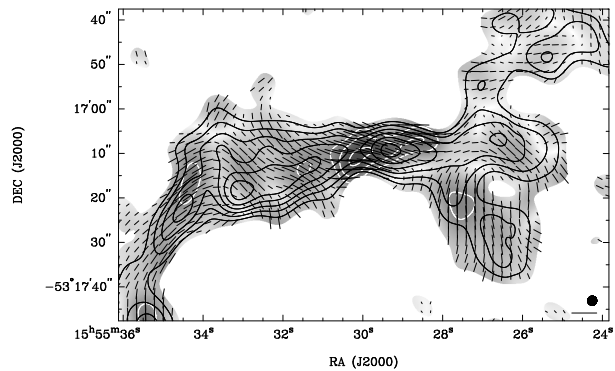}}
  \caption{Enlarged view of the central bar. The grey-scale and the white
contours show the linearly polarized intensity. The grey scale runs from
0 to 16 mJy\,beam$^{-1}$, the contours are at 7, 9 and 11 mJy\,beam$^{-1}$.
The dark contours show total intensity from 3 to 18 mJy\,beam$^{-1}$
in steps of 3 mJy\,beam$^{-1}$.
The orientation of the magnetic field is shown by the vector direction
with the length proportional to the linearly polarized intensity. The
beam is indicated in the bottom right corner, along with a vector of
length 11 mJy\,beam$^{-1}$.}
\end{figure}
Observationally, young SNRs such as Tycho and Kepler show
radial magnetic fields \cite{mil87}. In older SNRs, the structure tends 
to be more patchy, but toroidal fields seem more to be the norm.
Gull (1973)\nocite{gul73} first proposed that the radial fields were
the results of Rayleigh-Taylor instabilities forming at the interface
between the expanding shell and the ambient medium.
Jun \& Norman (1996)\nocite{jn96} carried out MHD simulations of 
expanding SNR shells and showed that the generally toroidal field
around the outer shell would be broken by radial `fingers' caused
by Rayleigh-Taylor instabilities. Although their simulation deals
with young SNRs, their results are very similar to the structure
of the magnetic field at the outer edges of SNR G328.4+0.2.
The magnetic field appears to be toroidal around the rim of the SNR
(especially in the north and west).  In the south-west there is some
indication of radial protusions poking through the rim.

Within the SNR, we see a frothy, bubble-like structure with the
magnetic fields tracing the total intensity in the filaments 
surrounding the bubbles. This is common in SNRs: in Vela, for example,
the magnetic field structure is highly correlated with
the filamentary structure \cite{mil95}.

MHD simulations of the evolution of a pulsar wind nebula have most
recently been performed by van der Swaluw (2003)\nocite{van03}.
He suggests that PWN should be elongated, with axial ratios of $\sim$2:1,
and that the magnetic field should run along the bar. Again, he deals
mainly with young SNRs, but the parallels with SNR G328.4+0.2 are clear.
The axial ratio of the bar is somewhat larger than 2:1 but the
magnetic field does indeed run along the major axis.
Van der Swaluw (2003) also speculates that if the reverse shock has had
time to `crush' the pulsar wind nebula, this would tend to increase
the axial ratio in the bar. It may be that in SNR G328.4+0.2 the
reverse shock is already interacting with the bar.

\subsection{Nature of the SNR and central pulsar}
Gaensler et al. (2000) put forward some convincing arguments 
for classifying the entire observable structure as a plerion, or
Crab-like SNR. However, the size and luminosity of G328.4+0.2 are
both considerably larger than the Crab, and the inferred properties
of the pulsar would make it the most energetic known.

We propose an alternative, equally plausible explanation which removes
the energetics problem. In our explanation,
the central bar structure is the pulsar wind
nebula and the rest of the SNR is the conventional synchrotron
emitting shock between the ejecta and the interstellar medium.
G328.4+0.2 would then be a composite SNR, with
properties more akin to G11.2--0.3 than the Crab.
Assuming a distance to the SNR of 17~kpc, the diameter of the SNR
is 25~pc and the extent of the bar is about $7\times$2~pc. The flux
density in the bar is only about one tenth of the entire SNR (as obtained
from the Gaensler et al. (2000) image).
Following the same line of reasoning as in Gaensler et al. (2000) we
determine the spin-down energy of the putative pulsar to be only
$3\times 10^{37}$~erg~s$^{-1}$ and an age for the SNR of $\sim$10~kyr.
This immediately removes the energetics problem and implies that the
pulsar has a current spin period of $\sim$50~ms and a standard magnetic
field strength of $2\times 10^{12}$~G. These parameters are very similar
to those of the pulsars in G11.2--0.3 ($P=65$~ms, $B=1.7\times 10^{12}$~G;
Torii et al. 1999\nocite{ttd+99}) and 3C58 ($P=66$~ms, $B=3.6\times 10^{12}$~G;
Murray et al. 2002\nocite{mss+02}). Indeed the radio structure
of G11.2--0.3 \cite{ggts88,trk02} and G328.4+0.2 are also similar, with 
a central asymmetric pulsar wind nebula surrounded by a `filled-in' shell
structure. In the case of G11.2--0.3 however, the wind nebula is much
more prominent in X-rays than in the radio.
In both G11.2--0.3 and 3C58, the pulsar is located at the centre
of the pulsar wind nebula rather than at one edge.
We surmise that the situation is likely
similar in G328.4+0.2. The geometric centre of the SNR is offset from
the centre of the bar structure which would require a pulsar of age 10~kyr
to have a velocity of 300 km~s$^{-1}$, well within the range of known
pulsar velocities.

\section{Summary}
Interferometric observations of SNR G328.4+0.2 at 19~GHz have been obtained
with the ATCA.
They show a highly polarized, circular SNR with a strong central bar.
The magnetic field lines are mostly toroidal at the edge of the SNR and run
parallel to the major axis of the bar. We propose that
this SNR is a composite remnant rather than a
Crab-like remnant as proposed by Gaensler et al. (2000) and there
are strong parallels between this SNR, G326.3--1.8 and G11.2--0.3.
High resolution X-ray observations are needed on this SNR. Non-thermal
emission is likely to be seen from the bar, but presumably with a smaller
extent due to the reduced synchrotron lifetimes in the X-ray. This will
give a clear indication of the current location of the pulsar,
which we believe should be at the centre of the bar.  With luck,
the pulsar itself may be detected through its pulsations as was the
case in 3C58.

\section*{Acknowledgments}
We acknowledge the excellence and the dedication of the 
members of the Receiver Group of the ATNF. 
The Australia Telescope is funded by the Commonwealth of 
Australia for operation as a National Facility managed by the CSIRO.
We thank Bryan Gaensler for providing his images and comments which
helped improve the text.

\bibliography{modrefs,psrrefs,crossrefs}
\bibliographystyle{mn}
\label{lastpage}
\end{document}